\documentclass[epj,english,twocolumn]{webofc}
\usepackage[varg]{txfonts}   
\usepackage{hyperref}        
\usepackage{tablefootnote}
\woctitle{Heavy Ion Accelerator Symposium 2015}
\begin{document}
\title{Time-dependent Hartree-Fock Study of Octupole Vibrations in doubly magic nuclei}
%
%

\author{C. Simenel\inst{1}\fnsep\thanks{\email{cedric.simenel@anu.edu.au}} \and
        J. Buete\inst{1} \and
        K. Vo-Phuoc\inst{1}}

\institute{Department of Nuclear Physics, RSPE, Australian National University, Canberra, Australian Capital Territory 2601, Australia
          }

\abstract{%
Octupole vibrations are studied in some doubly magic nuclei using the time-dependent Hartree-Fock (TDHF) theory with a Skyrme energy density functional. 
Through the use of the linear response theory, the energies and transition amplitudes of the low-lying vibrational modes for each of the nuclei were determined. 
Energies were found to be close to experimental results. 
However, transition amplitudes, quantified by the deformation parameter $\beta_3$, are underestimated by TDHF.
A comparison with single-particle excitations on the Hartree-Fock ground-state shows that the collective octupole vibrations have their energy lowered due to attractive RPA residual interaction.
}
\maketitle
\section{Introduction}
\label{intro}

\begin{figure}[!tb]
\centering
\includegraphics[width=3cm]{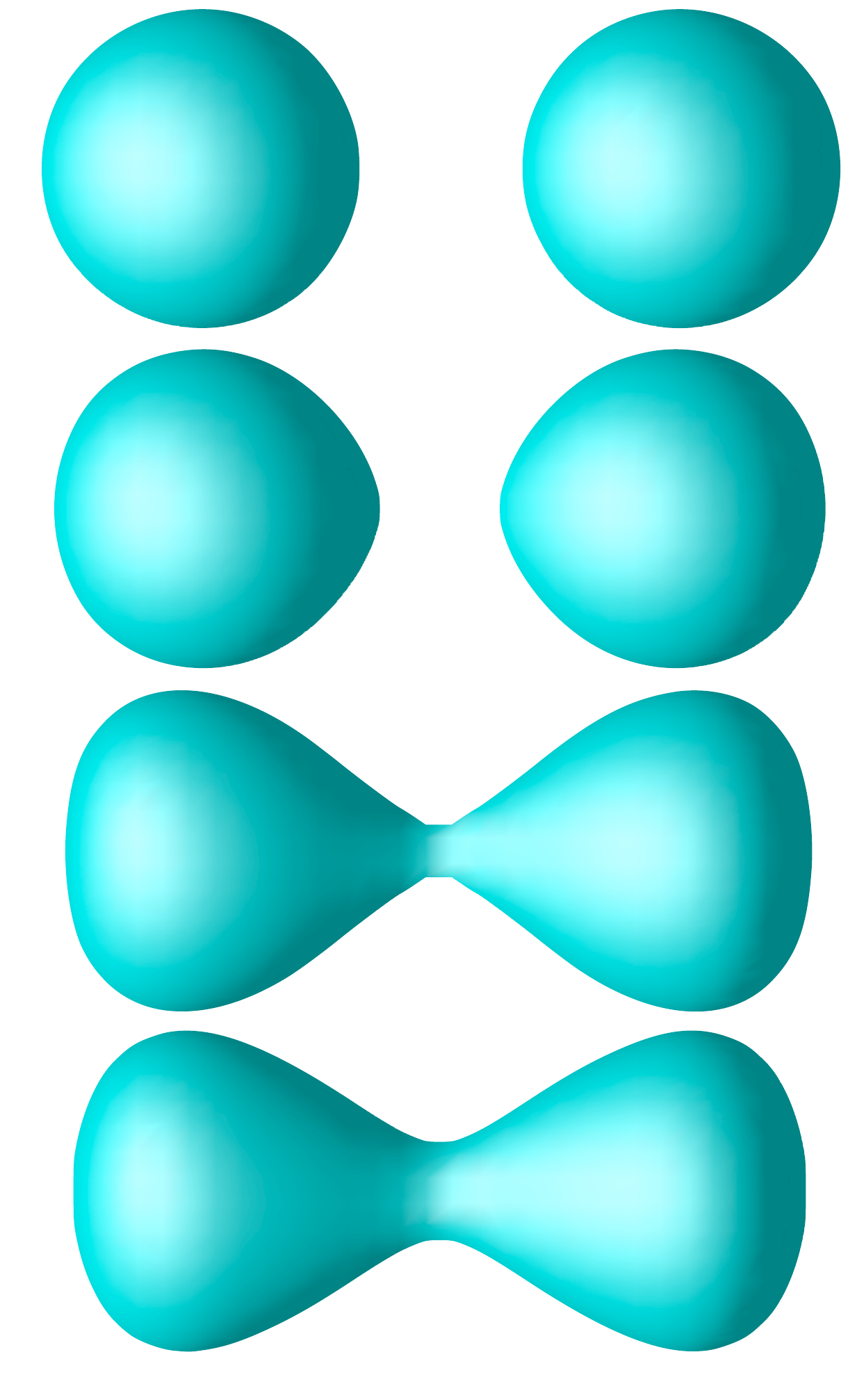}
\caption{TDHF isodensity surfaces at half the saturation density, $\rho_0/2=0.08$~fm$^{-3}$, for a $^{40}$Ca$+^{40}$Ca central collision at $E_{c.m.} = 53.3$~MeV. The time interval between each image is 0.25~zs.}
\label{fig:CaCa}       
\end{figure}

Nuclei can take all sorts of shapes: sphere, rugby ball, discus, pear, and (in theory) banana \cite{cha91}.
They can also oscillate between these shapes, leading to nuclear vibrations. 
The description of these vibrations is a fundamental test of quantum many-body models. 
In addition, vibrations strongly affect nuclear reactions \cite{das98} as shown by coupled-channel calculations \cite{hag12}. 
This has also been demonstrated with microscopic theory in the case of fusion \cite{sim01,sim07,obe12,sim13b,sim13c} 
 and fission \cite{sim14,sca15,god15,tan15,uma15a}. 
For instance, couplings between relative motion and low-lying octupole states induce the formation of a neck, as illustrated in Fig.~\ref{fig:CaCa}, which lowers the Coulomb barrier \cite{sim13c}. 
Reaction models should then incorporate a proper description of these vibrations in order to get a deep insight into the reaction mechanisms. 

Our basic tool is the time-dependent Hartree-Fock (TDHF) mean-field theory proposed by Dirac \cite{dir30}.
The TDHF approach describes the time-dependent motion of each nucleon wave-function, assuming that each nucleon moves freely in a mean-field generated by all the other nucleons. 
First TDHF applications to study the dynamics of nuclear systems already showed the ability of the approach to describe collective vibrations \cite{bon76,blo79} (see also \cite{neg82,sim12b} for reviews). 
Extensions including pairing correlations \cite{ave08,eba10,ste11,sca13b,sca14} as well as other components of the residual interaction \cite{toh95,lac01,toh02a,lac04} have also been used to describe nuclear vibrations in a dynamical way.

Several modern TDHF codes using realistic Skyrme energy density functionals are now available \cite{kim97,uma05,nak05,mar05}.
In order to study nuclear vibrations, these codes are used to compute the time evolution of various multipole moments after a small excitation of the nucleus. 
The linear response theory is then used to extract the energies and transition strengths of the first phonon of these vibrations. 
This approach allows for the study of both low-lying collective states and giant resonances, including their direct decay via nucleon emission \cite{cho87,ave13}.

In the small amplitude (linear) regime, the TDHF theory is fully equivalent to the RPA \cite{rin80} which is a standard tool to study collective vibrations.
Nevertheless, TDHF can also be used to investigate non-linear effects beyond the RPA \cite{sim03,rei07,sim09}.
Although the approach is able to describe both low-lying and high-lying vibrations, applications have mostly focussed on giant resonances studies \cite{blo79,cho87,lac00,ste04,alm05,nak05,uma05,mar05,rei06,ste07,ave13,sca13b,sca14}.
Despite their importance in understanding reaction mechanisms, only few applications have been made to investigate low-lying vibrations \cite{ave08,sca13b}.

In this contribution, we study low-lying octupole vibrations in doubly-magic nuclei. 
Indeed, low-lying octupole modes are known to strongly couple to the relative motion and thus to affect the outcome of the reactions \cite{mor94,lei95,hag97,row10}. 
Note that low-lying quadrupole phonons are also known to affect fusion \cite{das98}. However, their specific study will be the subject of a future work (see also \cite{sca13b}).

	\section{TDHF  simulations}

Pairing correlations are absent in mean-field calculations of doubly-magic nuclei due to large gaps at the Fermi level. 
The nucleus is initially in the Hartree-Fock (HF) ground-state computed with the \textsc{ev8} code \cite{bon05}.
The interaction between the nucleons is determined from the SLy4 parametrisation of the Skyrme functional \cite{cha98}.
The dynamics is computed with the \textsc{tdhf3d} code \cite{kim97} using the same functional. 
Details of the method can be found in Refs.~\cite{sim12b,sim13c}.

The vibrational modes are excited at the initial time with the  operator $\varepsilon\hat{Q}_{30}$ (we drop the indices as there is no ambiguity with other multipole operators here), where 
	\begin{equation}
		\hat{Q}_{30}\equiv\hat{Q} = \sqrt{\frac{7}{16\pi}}\sum^A_{i=1}[2\hat{x}^3 - 3\hat{x}(\hat{y}^2 + \hat{z}^2)] \label{octupole}
	\end{equation}
is the octupole moment operator, $A$ is the total number of nucleons and $\varepsilon$ is the boost intensity. 
The latter is small enough to be in the linear regime, i.e., the induced moment $\langle \hat{Q} \rangle(t)$ is proportional to $\varepsilon$. 

\begin{figure}[!tb]
\centering
\includegraphics[width=8.5cm]{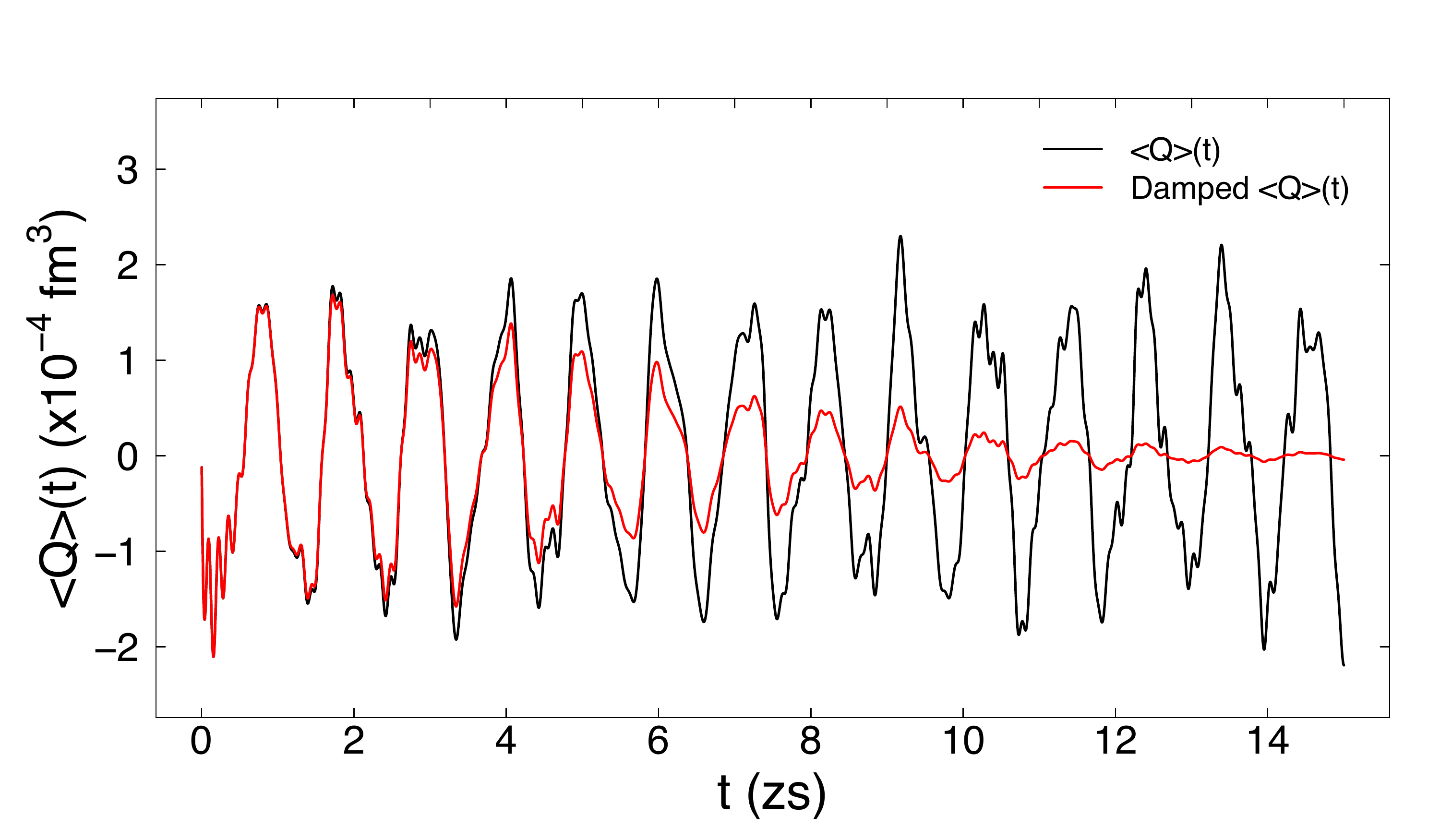}
\caption{Time evolution of the octupole moment in $^{40}$Ca following an octupole boost in the linear regime (black line).
The red line shows the same moment multiplied by a damping function. }
\label{fig:Q40Ca}       
\end{figure}

The calculations are performed on a three-dimensional grid with a plane of symmetry and a mesh spacing 0.8~fm. 
The calculations were run for 10000 iterations with time step $1.5\times10^{-24}$~s.
As an example, the time evolution of the octupole moment in $^{40}$Ca following an octupole boost at the initial time is shown in Fig.~\ref{fig:Q40Ca} (black line).
A strong low-frequency oscillation is clearly observed. 

\begin{figure}[!tb]
\centering
\includegraphics[width=8.5cm]{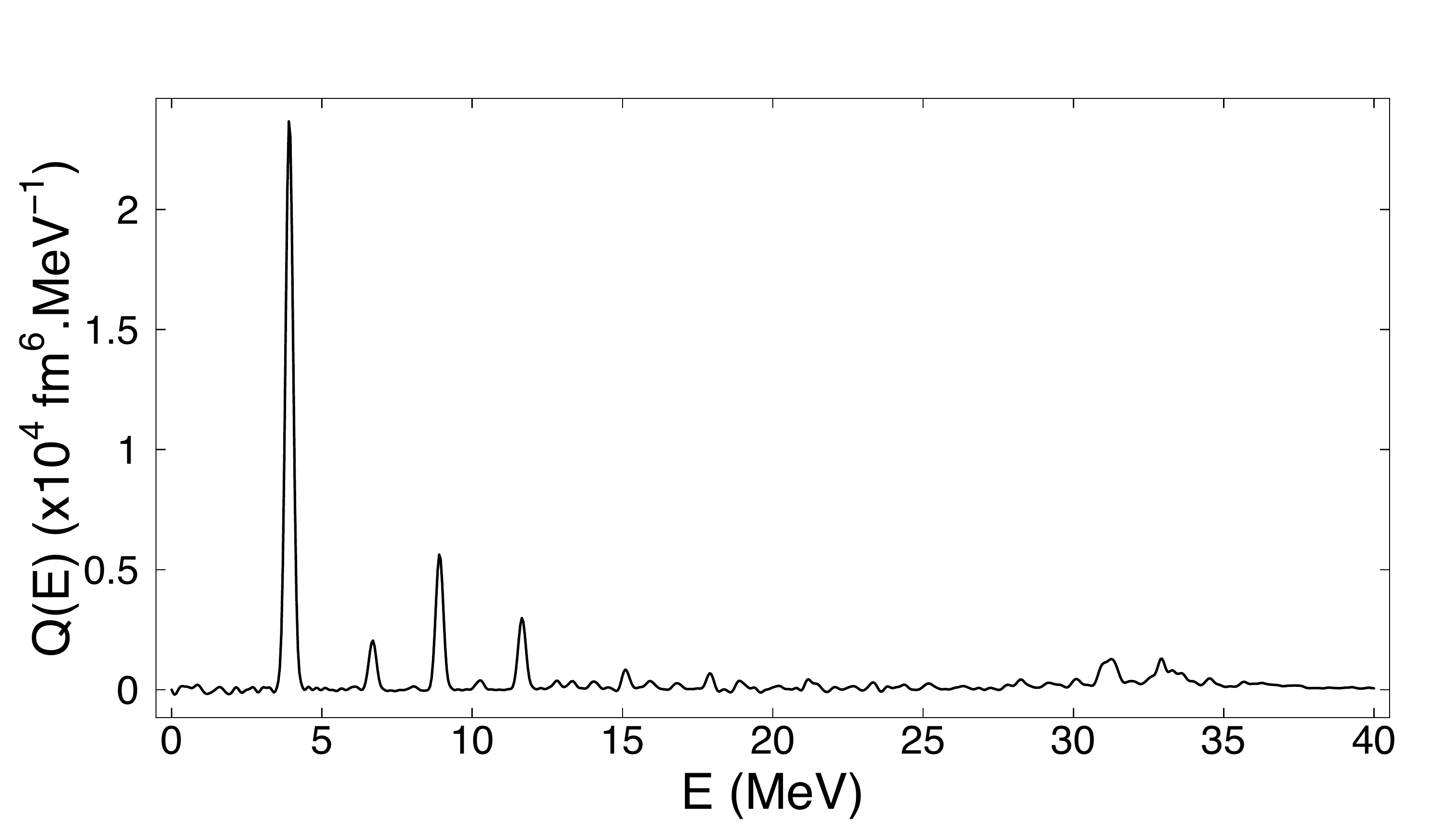}
\caption{Strength function obtained by Fourier transform of the damped octupole moment in Fig.~\ref{fig:Q40Ca}. }
\label{fig:S40Ca}       
\end{figure}

The strength function is obtained from a Fourier transform of  $\langle \hat{Q} \rangle(t)$.
However, to avoid spurious oscillations due to the finite time of the TDHF calculation, the octupole moment is multiplied by a damping function so that the oscillations disappear at the final time (red line in Fig.~\ref{fig:Q40Ca}). 
The resulting strength function is shown in Fig.~\ref{fig:S40Ca}. 
The main peak at $\sim3.9$~MeV corresponds to the first $3^-$ in $^{40}$Ca. 
The latter is found experimentally at $\sim3.7$~MeV \cite{kib02}, in good agreement with the calculations. 

It is interesting to compare the energy of the first phonon of a vibration with the individual energies of the particle-hole excitations the phonon is built on, i.e., with a coupling to $J^\pi=3^-$.
The latter are reported in Tab.~\ref{tab-1} for $^{40}$Ca. 
All particle-hole excitation energies are significantly larger than the energy of the $3^-_1$ phonon. 
This shows that the RPA residual interaction is attractive for the corresponding octupole vibration. 

\begin{table}[!t]
\centering
\caption{Energies of particle-hole excitations built on the $^{40}$Ca HF ground-state and coupling to $J^\pi=3^-$.}
\label{tab-1}       
		\begin{tabular}{cl}\hline
		
Configuration & Energy (MeV) \\\hline
			$\pi(1f_\frac{7}{2},1d_\frac{3}{2}^{-1})$ & 5.51 \\
			$\pi(1f_\frac{7}{2},2s_\frac{1}{2}^{-1})$ & 7.41 \\
			$\pi(1f_\frac{7}{2},1d_\frac{5}{2}^{-1})$ & 12.24\\
			$\nu(1f_\frac{7}{2},1d_\frac{3}{2}^{-1})$ & 5.69 \\
			$\nu(1f_\frac{7}{2},2s_\frac{1}{2}^{-1})$ & 7.74 \\
			$\nu(1f_\frac{7}{2},1d_\frac{5}{2}^{-1})$ & 12.60 \\\hline
		\end{tabular}
	\end{table}

In addition to the energy of the first phonon, the transition amplitude $q_3=\langle 3^-_1|\hat{Q}|0^+_1\rangle$ between the $0^+$ ground-state and the $3^-_1$ state can be obtained from the area of the corresponding peak in the strength function which is equal to $|q_3|^2$, i.e., the electric octupole transition probability $B(E3)\uparrow$. Note that the latter is usually measured experimentally for proton densities only \cite{kib02} while here it is associated to all nucleons. 
Usually, this transition probability is quantified via the deformation parameter $\beta_3$ which is an intermediate model-dependent quantity often used in coupled-channel calculations \cite{hag12}.
Its value can be determined according to \cite{kib02}
	\begin{equation}
		\beta_3 = \frac{4\pi q_3}{3 R_0^3 A},	\label{deformation}
	\end{equation}
where $R_0$ is the nuclear radius.
The latter is computed from the HF ground state using the isodensity at $\rho_0/2=0.08$~fm$^{-3}$.
The deformation parameter of the $3^-_1$ state in $^{40}$Ca is $\beta_3^{th.}\simeq0.20$ from the present TDHF calculations. 
Despite large  fluctuations in the experimental value $\beta_3^{exp.}\simeq 0.37-0.79$ which depends on the experimental probe \cite{kib02}, it remains that the TDHF prediction underestimates the data. 
Note that if we use the Fourier transform of the undamped $Q(t)$ to compute the area of the peak, the resulting value of $\beta_3$ increases by $\sim15\%$.

\begin{table}[!t]
\centering
\caption{Theoretical and experimental energy (in MeV) and deformation parameter of $3^-_1$ states in doubly-magic nuclei. Unless specified, the theoretical results are from  this work. Experimental data are from \cite{kib02} except for $^{132}$Sn (ENSDF).}
\label{tab-2}       
		\begin{tabular}{llllll}\hline
		Nucleus  & Ref. &  $E_3^{th.}$ & $\beta_3^{th.}$ & $E_3^{exp.}$ & $\beta_3^{exp.}$  \\\hline
		${}^{16}$O&  & 7.23 &0.30&  6.13 &0.37--0.79 \\
		${}^{40}$Ca&  & 3.92 &0.20&  3.74 &0.30--0.41 \\
		${}^{48}$Ca  &  &5.74 & 0.11 &4.51&  0.25\\
		${}^{56}$Ni  & & 9.59&0.13 & (4.93) &  - \\
		${}^{100}$Sn &  &6.91 & 0.08 & - &  - \\
		${}^{132}$Sn &  &5.53 & 0.07 &(4.35) &  - \\
		${}^{208}$Pb &\cite{sim12b}& 3.4 & 0.05 & 2.61   & 0.11 \\\hline
		\end{tabular}
	\end{table}

The energies and deformation parameters of the first octupole phonon are reported in 
 Tab. \ref{tab-2} for some doubly-magic nuclei. 
Overall, the calculated energies are in reasonable agreement with experimental data, except for $^{56}$Ni. 
However the experimental assignment to a $3^-$ state is uncertain. 
The origin of the large value of the $3^-$ energy in $^{56}$Ni is due to the complete filling of the $1f_{7/2}$ proton and neutron shells. 
Then, the only possibility to produce a $3^-$ is to promote a $1f_{7/2}$ nucleon to the high-lying $1g_{9/2}$ shell, or to excite a nucleon from below the $Z,N=20$ magic gap to the $1d$ shells. 
Both types of excitations require a large amount of energy, as shown in Tab.~\ref{tab-3}.

\begin{table}[!t]
\centering
\caption{Same as Tab.~\ref{tab-2} for $^{56}$Ni.}
\label{tab-3}       
		\begin{tabular}{cl}\hline
Configuration & Energy (MeV) \\\hline
 $\pi(1g_\frac{9}{2},1f_\frac{7}{2}^{-1})$ &11.761 \\
 $\pi(2p_\frac{3}{2},1d_\frac{3}{2}^{-1})$ & 11.768\\
 $ \pi(1f_\frac{5}{2},1d_\frac{3}{2}^{-1})$ & 15.080\\
 $\nu(1g_\frac{9}{2},1f_\frac{7}{2}^{-1})$ &12.181 \\
 $\nu(2p_\frac{3}{2},1d_\frac{3}{2}^{-1})$ & 12.262\\
$\nu(1f_\frac{5}{2},1d_\frac{3}{2}^{-1})$ & 15.710\\\hline
		\end{tabular}
	\end{table}

Finally, the underestimation of the deformation parameter, already noted in $^{40}$Ca, seems systematic. 
It would be interesting to investigate the origin of this effect which could come from a limitation of the mean-field approximation, from the choice of the  interaction, or from the way this parameter is deduced experimentally. 

\section{Conclusions}

Octupole vibrations in doubly-magic nuclei have been studied with TDHF and the linear response theory.
The energy of the first phonons are in relatively good agreement with experimental data. 
In particular, variations induced by the underlying single-particle shell structure are well captured.

The deformation parameters, however, are underestimated in the present calculations. 
The origin of this discrepancy is unclear and deserves further studies.
In particular, other parametrisations of the Skyrme functional should be tested.

Nevertheless, the TDHF results capture reasonably well the dependence of the phonon characteristics (energy and deformation parameter) with the mass of the nucleus. 
This indicates that TDHF is indeed a good tool in order to investigate the interplay between nuclear structure and reaction mechanisms. 
For instance, it was shown in Ref.~\cite{sim13c} that, unlike $^{56}$Ni, $^{40}$Ca nuclei could develop octupole shapes in collision with a partner thanks to their low-lying collective $3^-$ state (see Fig.~\ref{fig:CaCa}). 

\acknowledgement

This work has been supported by the Australian Research Council under Grant No. FT120100760.
This research was undertaken on the NCI National Facility in Canberra, Australia, which is supported by the Australian Commonwealth Government.


%
 \bibliography{biblio}

\begin{thebibliography}{49}

\bibitem{cha91}
R.~Chasman, Phys. Lett. B \textbf{266}, 243 (1991)

\bibitem{das98}
M.~Dasgupta, D.J. Hinde, N.~Rowley, A.M. Stefanini, Ann. Rev. Nucl. Part. Sci.
  \textbf{48}, 401 (1998)

\bibitem{hag12}
K.~Hagino, N.~Takigawa, Prog. Th. Phys. \textbf{128}, 1061 (2012)

\bibitem{sim01}
C.~Simenel, P.~Chomaz, G.~de~France, Phys. Rev. Lett. \textbf{86}, 2971 (2001)

\bibitem{sim07}
C.~Simenel, P.~Chomaz, G.~de~France, Phys. Rev. C \textbf{76}, 024609 (2007)

\bibitem{obe12}
V.E. Oberacker, A.S. Umar, J.A. Maruhn, P.G. Reinhard, Phys. Rev. C
  \textbf{85}, 034609 (2012)

\bibitem{sim13b}
C.~Simenel, R.~Keser, A.S. Umar, V.E. Oberacker, Phys. Rev. C \textbf{88},
  024617 (2013)

\bibitem{sim13c}
C.~Simenel, M.~Dasgupta, D.J. Hinde, E.~Williams, Phys. Rev. C \textbf{88},
  064604 (2013)

\bibitem{sim14}
C.~Simenel, A.S. Umar, Phys. Rev. C \textbf{89}, 031601 (2014)

\bibitem{sca15}
G.~Scamps, C.~Simenel, D.~Lacroix, Phys. Rev. C \textbf{92}, 011602 (2015)

\bibitem{god15}
P.~Goddard, P.~Stevenson, A.~Rios, Phys. Rev. C \textbf{92}, 054610 (2015)

\bibitem{tan15}
Y.~Tanimura, D.~Lacroix, G.~Scamps, Phys. Rev. C \textbf{92}, 034601 (2015)

\bibitem{uma15a}
A.S. Umar, V.E. Oberacker, C.~Simenel, Phys. Rev. C \textbf{92}, 024621 (2015)

\bibitem{dir30}
P.A.M. Dirac, Proc. Camb. Phil. Soc. \textbf{26}, 376 (1930)

\bibitem{bon76}
P.~Bonche, S.~Koonin, J.W. Negele, Phys. Rev. C \textbf{13}, 1226 (1976)

\bibitem{blo79}
J.~B{\l}ocki, H.~Flocard, Phys. Lett. B \textbf{85}, 163 (1979)

\bibitem{neg82}
J.W. Negele, Rev. Mod. Phys. \textbf{54}, 913 (1982)

\bibitem{sim12b}
C.~Simenel, Eur. Phys. J. A \textbf{48}, 152 (2012)

\bibitem{ave08}
B.~Avez, C.~Simenel, P.~Chomaz, Phys. Rev. C \textbf{78}, 044318 (2008)

\bibitem{eba10}
S.~Ebata, T.~Nakatsukasa, T.~Inakura, K.~Yoshida, Y.~Hashimoto, K.~Yabana,
  Phys. Rev. C \textbf{82}, 034306 (2010)

\bibitem{ste11}
I.~Stetcu, A.~Bulgac, P.~Magierski, K.J. Roche, Phys. Rev. C \textbf{84},
  051309 (2011)

\bibitem{sca13b}
G.~Scamps, D.~Lacroix, Phys. Rev. C \textbf{88}, 044310 (2013)

\bibitem{sca14}
G.~Scamps, D.~Lacroix, Phys. Rev. C \textbf{89}, 034314 (2014)

\bibitem{toh95}
M.~Tohyama, Prog. Theor. Phys. \textbf{94}, 147 (1995)

\bibitem{lac01}
D.~Lacroix, S.~Ayik, P.~Chomaz, Phys. Rev. C \textbf{63}, 064305 (2001)

\bibitem{toh02a}
M.~Tohyama, A.S. Umar, Phys. Lett. B \textbf{549}, 72 (2002)

\bibitem{lac04}
D.~Lacroix, S.~Ayik, P.~Chomaz, Prog. Part. Nucl. Phys. \textbf{52}, 497 (2004)

\bibitem{kim97}
K.H. Kim, T.~Otsuka, P.~Bonche, J. Phys. G \textbf{23}, 1267 (1997)

\bibitem{uma05}
A.S. Umar, V.E. Oberacker, Phys. Rev. C \textbf{71}, 034314 (2005)

\bibitem{nak05}
T.~Nakatsukasa, K.~Yabana, Phys. Rev. C \textbf{71}, 024301 (2005)

\bibitem{mar05}
J.A. Maruhn, P.G. Reinhard, P.D. Stevenson, J.R. Stone, M.R. Strayer, Phys.
  Rev. C \textbf{71}, 064328 (2005)

\bibitem{cho87}
P.~Chomaz, N.V. Giai, S.~Stringari, Phys. Lett. B \textbf{189}, 375 (1987)

\bibitem{ave13}
B.~Avez, C.~Simenel, Eur. Phys. J. A \textbf{49}, 76 (2013)

\bibitem{rin80}
P.~Ring, P.~Schuck, \emph{The Nuclear Many-Body Problem} (Springer Verlag,
  1980)

\bibitem{sim03}
C.~Simenel, P.~Chomaz, Phys. Rev. C \textbf{68}, 024302 (2003)

\bibitem{rei07}
P.G. Reinhard, L.~Guo, J.A. Maruhn, Eur. Phys. J. A \textbf{32}, 19 (2007)

\bibitem{sim09}
C.~Simenel, P.~Chomaz, Phys. Rev. C \textbf{80}, 064309 (2009)

\bibitem{lac00}
D.~Lacroix, A.~Mai, P.~von Neumann-Cosel, A.~Richter, J.~Wambach, Phys. Lett. B
  \textbf{479}, 15 (2000)

\bibitem{ste04}
P.D. Stevenson, M.R. Strayer, J.~Rikovska~Stone, W.G. Newton, Int. J. Mod.
  Phys. E \textbf{13}, 181 (2004)

\bibitem{alm05}
D.~Almehed, P.D. Stevenson, J. Phys. G \textbf{31}, S1819 (2005)

\bibitem{rei06}
P.G. Reinhard, P.D. Stevenson, D.~Almehed, J.A. Maruhn, M.R. Strayer, Phys.
  Rev. E \textbf{73}, 036709 (2006)

\bibitem{ste07}
P.D. Stevenson, D.~Almehed, P.G. Reinhard, J.A. Maruhn, Nucl. Phys. A
  \textbf{788}, 343 (2007)

\bibitem{mor94}
C.R. Morton, M.~Dasgupta, D.J. Hinde, J.R. Leigh, R.C. Lemmon, J.P. Lestone,
  J.C. Mein, J.O. Newton, H.~Timmers, N.~Rowley et~al., Phys. Rev. Lett.
  \textbf{72}, 4074 (1994)

\bibitem{lei95}
J.R. Leigh, M.~Dasgupta, D.J. Hinde, J.C. Mein, C.R. Morton, R.C. Lemmon, J.P.
  Lestone, J.O. Newton, H.~Timmers, J.X. Wei et~al., Phys. Rev. C \textbf{52},
  3151 (1995)

\bibitem{hag97}
K.~Hagino, N.~Takigawa, M.~Dasgupta, D.J. Hinde, J.R. Leigh, Phys. Rev. Lett.
  \textbf{79}, 2014 (1997)

\bibitem{row10}
N.~Rowley, K.~Hagino, Nucl. Phys. A \textbf{834}, 110c  (2010), the 10th
  International Conference on Nucleus-Nucleus Collisions (NN2009)

\bibitem{bon05}
P.~Bonche, H.~Flocard, P.H. Heenen, Comp. Phys. Com. \textbf{171}, 49 (2005)

\bibitem{cha98}
E.~Chabanat, P.~Bonche, P.~Haensel, J.~Meyer, R.~Schaeffer, Nucl. Phys. A
  \textbf{635}, 231 (1998)

\bibitem{kib02}
T.~Kib\'edi, R.H. Spear, At. Dat. Nucl. Dat. Tab. \textbf{80}, 35  (2002)

\end{thebibliography}
%
%
%
%

\end{document}